\documentstyle[11pt]{article}
\begin{document}
\title{MEAN CURVATURE DRIVEN RICCI FLOW}

\author{{\bf Victor Tapia}\\
\\
{\it Departamento de Matem\'aticas}\\
{\it Universidad Nacional de Colombia}\\
{\it Bogot\'a, Colombia}\\
\\
{\tt tapiens@gmail.com}}

\maketitle

\begin{abstract}

We obtain the evolution equations for the Riemann tensor, the Ricci tensor and the scalar curvature induced by the mean curvature flow. The evolution for the scalar curvature is similar to the Ricci flow, however, negative, rather than positive, curvature is preserved. Our results are valid in any dimension.

\end{abstract}

\section{Introduction}

One of the most interesting problems in geometric analysis is the description of the evolution of Riemannian spaces under particular flows. The best known examples are the Ricci flow and the mean curvature flow. The Ricci flow \cite{Ha82} describes how a three--dimensional Riemannian space evolves towards a highly symmetric, homogeneous, configuration. The application of the Ricci flow in the demostration of the Thurston geometrisation conjecture \cite{Th82} and, consequently, the resolution of the Poincar\'e conjecture is well--known \cite{Pe02,Pe03a,Pe03b}. On the other hand, the mean curvature flow \cite{Hu84} describes the evolution of surfaces as embedded in a higher dimensional, possibly flat, space. In this case the relevant equation is a heat--like equation foe the functions describing the embedding.

Due to a series of remarkable theorems, \cite{Ja26,Ca27,Bu31,Fr61,Na56,Cl70,Gr70} any Riemannian space can be considered as a space embedded into a higher--dimensional flat space. Therefore, the extrinsic approach to Riemannian geometry is equivalent to the usual intrinsic approach to Riemannian geometry in which no reference to an embedding space is done. In the intrinsic approach to Riemannian geometry the metric tensor is the basic geometrical object, while in the extrinsic approach the basic geometrical objects are the embedding functions describing the situation of the embedded space as a subspace of the ambient space. 

The Ricci flow is based on intrinsic Riemannian geometry while the mean curvature flow is based on extrinsic Riemannian geometry. Due to the embedding theorems both approaches to Riemannian geometry are equivalent. Therefore, the Ricci flow can be interpreted extrinsically. In fact, in several approaches Riemannian spaces are pictured as embedded spaces, even when their evolution is given in terms of the metric tensor rather than in terms of the embedding functions. Therefore, the evolution of these spaces, even when pictured extrinsically, is intrinsic, that is, the one given by the evolution of the metric tensor. On the other hand, the mean curvature flow can be interpreted intrinsically.

The purpose of this work is to obtain the evolution equations for the elements of intrinsic Riemannian geometry (metric tensor, Riemann  tensor, Ricci tensor, scalar curvature) induced by the mean curvature flow. The metric tensor evolves according to an evolution equation which is driven by the Gauss tensor of the embedding. The scalar curvature evolves according to an equation similar to the Ricci flow and therefore has similar properties, that is, spaces evolve towards highly symmetric, homogeneous, configurations. The main difference with respect to the Ricci flow is that, due to the presence of a minus sign, negative, rather than positive, curvature is preserved.

Sections 2 and 3 are preliminaries for Riemannian geometry and geometric flows, respectively. Section 4 contains our main results, namely, the evolution equations for the curvature tensors. Section 5 contains the conclusions.

\section{Riemannian geometry}

Let us start with some fundamentals of Riemannian geometry \cite{Sc54}. The basic geometrical object in Riemannian geometry is the metric tensor ${\bf g}$ with components $g_{ij}({\bf x})$. The Christoffel symbol is given by

\begin{equation}
{\Gamma^k}_{ij}={1\over2}\,g^{k\ell}\,\left[{{\partial g_{j\ell}}\over{\partial x^i}}+{{\partial g_{i\ell}}\over{\partial x^j}}-{{\partial g_{ij}}\over{\partial x^\ell}}\right]\,,
\label{01}
\end{equation}

\noindent and the covariant derivative of a covariant vector is given by

\begin{equation}
\nabla_i v_j={{\partial v_j}\over{\partial x^i}}-{\Gamma^k}_{ij}\,v_k\,.
\label{02}
\end{equation}

\noindent The covariant derivative extends in a natural way to contravariant vectors and tensors of any covariance so as to preserve the Leibniz rule.

The Riemann tensor is given by

\begin{equation}
{R^k}_{\ell ij}={{\partial{\Gamma^k}_{j\ell}}\over{\partial x^i}}-{{\partial{\Gamma^k}_{i\ell}}
\over{\partial x^j}}+{\Gamma^k}_{im}\,{\Gamma^m}_{j\ell}-{\Gamma^k}_{jm}\,{\Gamma^m}_{i\ell}\,.
\label{03}
\end{equation}

\noindent Covariant derivatives do not commute, and we obtain the Ricci identity for a covariant vector

\begin{equation}
\nabla_i\nabla_j v_k-\nabla_j\nabla_i v_k=-{R^\ell}_{kij}\,v_\ell\,.
\label{04}
\end{equation}

\noindent The Ricci identity extends in a natural way to contravariant vectors and tensors of any covariance.

The completely covariant Riemann tensor, $R_{ijk\ell}=g_{im}{R^m}_{jk\ell}$, is given by

\begin{eqnarray}
R_{ijk\ell}({\bf g})&=&{1\over2}\,\left[{{\partial^2g_{jk}}\over{\partial x^i\partial x^\ell}}+{{
\partial^2g_{i\ell}}\over{\partial x^j\partial x^k}}-{{\partial^2g_{j\ell}}\over{\partial x^i
\partial x^k}}-{{\partial^2g_{ik}}\over{\partial x^j\partial x^\ell}}\right]\nonumber\\
&&+g_{mn}\,\left[{
\Gamma^m}_{i\ell}\,{\Gamma^n}_{jk}-{\Gamma^m}_{ik}\,{\Gamma^n}_{j\ell}\right]\,.
\label{05}
\end{eqnarray}

\noindent The Ricci tensor and the scalar curvature are given by the usual expressions

\begin{eqnarray}
R_{j\ell}({\bf g})&=&g^{ik}\,R_{ijk\ell}({\bf g})\,,\label{06}\\
R({\bf g})&=&g^{j\ell}\,R_{j\ell}({\bf g})\,.
\label{07}
\end{eqnarray}

\noindent Finally, the completely covariant Riemann tensor (\ref{05}) satisfies the Bianchi identity

\begin{equation}
\nabla_i R_{jk\ell m}({\bf g})+\nabla_j R_{ki\ell m}({\bf g})+\nabla_k R_{ij\ell m}({\bf g})
\equiv0\,.
\label{08}
\end{equation}

Let us now consider the extrinsic approach to Riemannian geometry. Let $V_n$ be a Riemannian space with metric tensor ${\bf g}$ with component $g_{ij}({\bf x})$. According to the local and global embedding theorems \cite{Ja26,Ca27,Bu31,Fr61,Na56,Cl70,Gr70} there always exist functions $X^A=X^A({\bf x})$, $A=1,\cdots,N$, and $G_{AB}$, $N\geq n$, with ${\rm Rie}({\bf G})=0$, such that the components of the metric tensor ${\bf g}$ can be written as

\begin{equation}
g_{ij}=G_{AB}\,{X^A}_i\,{X^B}_j\,,
\label{11}
\end{equation}

\noindent where ${X^A}_i=\partial X^A/\partial x^i$. Therefore, any Riemannian space $V_n$ can be considered as a space embedded in a higher--dimensional flat space $E_N$, $N\geq n$, with metric tensor ${\bf G}$ with components $G_{AB}$ in local coordinates $X^A$. Then, $g_{ij}$ as given by (\ref{11}) is the induced metric tensor. For later convenience we choose Minkowskian coordinates in $E_N$ such that $G_{AB}=\eta_{AB}={\rm diag}(\pm1,\cdots,\pm1)$. Denoting the coordinates $X^A$ by ${\bf X}$ we rewrite (\ref{11}) as

\begin{equation}
g_{ij}={\bf X}_i\,\cdot\,{\bf X}_j\,,
\label{12}
\end{equation}

\noindent where ${\bf X}_i=\partial{\bf X}/\partial x^i$ and the dot `$\cdot$' denotes the inner product with $\eta_{AB}$.

Now we remind some elementary results in extrinsic Riemannian geometry \cite{Ei26}. An important object in extrinsic Riemannian geometry is the Gauss tensor, which is given by

\begin{equation}
\nabla_{ij}{\bf X}={\bf X}_{ij}-{\Gamma^k}_{ij}\,{\bf X}_k\,.
\label{13}
\end{equation}

\noindent where ${\bf X}_{ij}=\partial^2{\bf X}/\partial x^i\partial x^j$ and $\nabla_{ij}=\nabla_i\nabla_j$ are second--order covariant derivatives.

The Gauss tensor satisfies the important identity

\begin{equation}
\nabla_{ij}{\bf X}\,\cdot\,\nabla_k{\bf X}\equiv0\,,
\label{14}
\end{equation}

\noindent where obviously $\nabla_k{\bf X}={\bf X}_k$. Identity (\ref{14}) is used extensively in the rest of this work. The completely covariant Riemann tensor (\ref{05}) is given by

\begin{equation}
R_{ijk\ell}=\nabla_{ik}{\bf X}\,\cdot\,\nabla_{j\ell}{\bf X}-\nabla_{i\ell}{\bf X}\,\cdot\,
\nabla_{jk}{\bf X}\,.
\label{15}
\end{equation}

\noindent This is the Gauss equation of the embedding.

\section{Geometric flows}

In {\tt 1982} Hamilton \cite{Ha82} introduced the Ricci flow, which is a differential equation generalizing some features of the heat equation for the components of a positive definite metric tensor. The Ricci flow is explicitly given by

\begin{equation}
\partial_t g_{ij}=-2\,R_{ij}({\bf g})\,.
\label{09}
\end{equation}

\noindent Due to the minus sign, spaces are contracted in the direction of positive Ricci curvature, while are expanded in the direction of negative Ricci curvature. The scalar curvature evolves according to

\begin{equation}
\partial_t R=\nabla^2R+2\,R_{ij}\,R^{ij}\,,
\label{10}
\end{equation}

\noindent where $\nabla^2=g^{ij}\nabla_{ij}$ is the Laplacian and $R_{ij}R^{ij}\geq0$. The Ricci flow makes a Riemannian space to evolve towards stationary solutions which are constant curvature configurations. Furthermore, spaces with positive curvature keep on having positive curvature.

One can also consider extrinsic flows closer in spirit to what one has in mind when picturing evolving surfaces. In the extrinsic approach to Riemannian geometry the basic geometrical objects are the embedding functions. Therefore, it would be natural to describe the evolution of embedded spaces in term of these functions. In analogy with the heat equation of thermodynamics one considers the mean curvature flow \cite{Hu84}

\begin{equation}
\partial_t{\bf X}=\nabla^2{\bf X}\,,
\label{16}
\end{equation}

\noindent where now the Laplacian is with respect to the induced metric tensor ${\bf g}$ as given by (\ref{12}). The stationary solution of equation (\ref{16}), $\partial_t{\bf X}=0$, is given by the condition $\nabla^2{\bf X}=0$, which corresponds to minimal surfaces \cite{ES64,Ta89}.

The mean curvature flow is well--known and has been widely studied in the literature. However, our forthcoming analysis is new. 

\section{Evolution of curvature tensors}

The evolution of intrinsic quantities under the mean curvature flow is obtained just by combining (\ref{16}) with the corresponding definitions, (\ref{12}) and (\ref{15}). For the metric tensor we obtain

\begin{equation}
\partial_t g_{ij}=-2\,\nabla^2{\bf X}\,\cdot\,\nabla_{ij}{\bf X}\,.
\label{17}
\end{equation}

\noindent Therefore, the evolution of the metric tensor is driven by the Gauss tensor (\ref{13}). The stationary solutions of equation (\ref{17}), $\partial_t g_{ij}=0$, are given by the condition $\nabla^2{\bf X}\cdot\nabla_{ij}{\bf X}=0$. Two possible solutions are: $\nabla^2{\bf X}=0$ which corresponds to minimal surfaces as above, and $\nabla_{ij}{\bf X}=0$, but this implies, according to (\ref{15}), a flat space.

Now we consider the evolution of the Riemann tensor, the Ricci tensor and the scalar curvature. The time derivative of the Riemann tensor can be written, using (\ref{05}), completely in terms of derivatives of the metric tensor. Then, replacing from (\ref{17}) we would obtain the evolution equation for the Riemann tensor induced by the mean curvature flow. However, we can also combine (\ref{13}) and (\ref{16}) which is the way in which we proceed now.

The time derivative of the Riemann tensor is given by

\begin{eqnarray}
\partial_t R_{ijk\ell}&=&\nabla_{ik}\nabla^2{\bf X}\,\cdot\,\nabla_{j\ell}{\bf X}+\nabla_{ik}{\bf X}\,\cdot\,\nabla_{j\ell}\nabla^2{\bf X}\nonumber\\
&&-\nabla_{i\ell}\nabla^2{\bf X}\,\cdot\,\nabla_{jk}{\bf X}-\nabla_{i\ell}{\bf X}\,\cdot\,\nabla_{jk}\nabla^2{\bf X}\,.
\label{18}
\end{eqnarray}

\noindent The right--hand side of equation (\ref{18}) can be rewritten in terms of second--order covariant derivatives of the Riemann and Ricci tensors. The only terms with the right symmetries which must be considered are

\begin{eqnarray}
A_{ijk\ell}&=&\nabla^2R_{ijk\ell}\,,\nonumber\\
B_{ijk\ell}&=&\nabla_{ik}R_{\ell j}-\nabla_{jk}R_{\ell i}-\nabla_{i\ell}R_{kj}+\nabla_{j\ell}
R_{ki}\nonumber\\
&&+\nabla_{ki}R_{j\ell}-\nabla_{\ell i}R_{jk}-\nabla_{kj}R_{i\ell}+\nabla_{\ell j}R_{ik}\,.
\label{19}
\end{eqnarray}

\noindent All other possible terms are reducible to these ones by means of the Bianchi identity (\ref{08}) and the Ricci identity (\ref{04}), and the differences are at most quadratic in the curvature tensor. After a lenghty but straightforward calculation we obtain

\begin{eqnarray}
2\,\partial_t R_{ijk\ell}&=&4\,\nabla^2R_{ijk\ell}-B_{ijk\ell}\nonumber\\
&&-4\,\left[R_{mjk\ell}\,{R^m}_i+R_{imk\ell}\,{R^m}_j+R_{ijm\ell}\,{R^m}_k+R_{ijkm}\,{R^m}_\ell
\right]\nonumber\\
&&+8\,{{{R^m}_k}^n}_i\,R_{mjn\ell}-8\,{{{R^m}_k}^n}_j\,R_{min\ell}+4\,{R^{mn}}_{ij}\,R_{mnk\ell}
\nonumber\\
&&+{R^m}_{jk\ell}\,\partial_t g_{im}-{R^m}_{ik\ell}\,\partial_t g_{jm}+{R^m}_{\ell ij}\,\partial_t g_{km}-{R^m}_{kij}\,\partial_t g_{\ell m}\,.\nonumber\\
&&
\label{20}
\end{eqnarray}

\noindent The time derivatives of the Ricci tensor and the scalar curvature are given by

\begin{eqnarray}
2\,\partial_t R_{j\ell}&=&2\,\nabla^2R_{j\ell}+2\,\left[\nabla_{mj}{R^m}_\ell+\nabla_{m\ell}
{R^m}_j\right]-2\,\nabla_{j\ell}R\nonumber\\
&&-8\,R_{mj}\,{R^m}_\ell-4\,R_{mnpj}\,{R^{mnp}}_\ell\nonumber\\
&&+\left[{R^m}_j\,\partial_t g_{\ell m}+{R^m}_\ell\,\partial_t g_{jm}\right]\,,
\label{21}\\
\partial_t R&=&\nabla^2R-4\,R_{ij}\,R^{ij}-2\,R_{ijk\ell}\,R^{ijk\ell}\,.
\label{22}
\end{eqnarray}

\noindent The last equation, (\ref{22}), is similar to (\ref{10}), except for the presence of minus signs in the last terms. For positive definite metric tensors we have $R_{ij}R^{ij}>0$ and $R_{ijk\ell}R^{ijk\ell}>0$. Therefore, if we consider $S=-R$ as our unknown function, then, we have that negative, rather than positive, curvature is preserved under this flow.

Equation (\ref{22}) is valid for any dimension. However, for two and three dimensions the Riemann tensor acquires particularly simple forms and equation (\ref{22}) reduces to

\begin{eqnarray}
\partial_t R^{(2)}&=&\nabla^2R^{(2)}-4\,[R^{(2)}]^2\,.
\label{23}\\
\partial_t R^{(3)}&=&\nabla^2R^{(3)}-12\,[{\rm Ric}^{(3)}]^2+2\,[R^{(3)}]^2\,.
\label{24}
\end{eqnarray}

\noindent In dimension $d\geq4$ we obtain

\begin{equation}
\partial_t R^{(d)}=\nabla^2R^{(d)}-{{4d}\over{(d-2)}}\,[{\rm Ric}^{(d)}]^2+{4\over{(d-2)(d-1)}}\,[R^{(d)}]^2-2\,[{\rm Weyl}^{(d)}]^2\,.
\label{25}
\end{equation}

\section{Conclusions}

In conclusion, the evolution induced by the mean curvature flow in the curvature tensor, has several interesting properties. Firstly, all developments are valid for any dimension and for any signature. Secondly, for positive definite metric tensors, the evolution equation for the scalar curvature, (\ref{22}), has the same generic properties as the Ricci flow. The evolution is towards constant curvature spaces and negative, rather than positive, curvature is preserved.

\section*{Acknowledgements}

The author thanks Fernando Zalamea for inviting him to deliver a seminar on geometric analysis where the idea of this work was born. Some of the calculations were kindly and patiently checked by Alexander Torres. The author thanks Mikhail Malakhaltsev for useful criticism. This work has been suported by Facultad de Ciencias, Universidad Nacional de Colombia, under contract 8180.


\end{document}